\documentclass[conference, comsoc]{IEEEtran}
\IEEEoverridecommandlockouts
\usepackage[left=0.625in,right=0.625in,top=0.75in,bottom=1in]{geometry}

\usepackage{caption}
\usepackage{subcaption}
\usepackage{multirow}
\usepackage{array}
\usepackage{balance}
\usepackage{cite}
\usepackage{amsmath,amssymb,amsfonts}
\usepackage{graphicx}
\usepackage{textcomp}
\usepackage{bm}
\usepackage{nicefrac}
\usepackage{mathtools}
\usepackage{comment}
\usepackage{pgfplots}
\usepackage{pgfkeys,pgfmath,pgfcore}
\pgfplotsset{compat=1.18}
\pgfkeys{/pgf/number format/.cd,fixed,precision=2}
\usepackage[ruled, vlined]{algorithm2e}
\usepackage{multirow}
\usepackage{xcolor} 
\usepackage{colortbl}
\definecolor{verylightgray}{rgb}{0.9,0.9,0.9}
\usepackage[colorlinks=true,allcolors=blue,bookmarks=false]{hyperref}
\def\BibTeX{{\rm B\kern-.05em{\sc i\kern-.025em b}\kern-.08em
    T\kern-.1667em\lower.7ex\hbox{E}\kern-.125emX}}
\DeclareMathOperator{\myvec}{vec}

\begin{document}

\title{Building 6G Radio Foundation Models \\with Transformer Architectures}

\author{\IEEEauthorblockN{ Ahmed Aboulfotouh\IEEEauthorrefmark{3}, Ashkan Eshaghbeigi\IEEEauthorrefmark{1}, and
Hatem Abou-Zeid\IEEEauthorrefmark{3}
}
\IEEEauthorblockA{\IEEEauthorrefmark{3}{Department of Electrical and Software Engineering}, 
{University of Calgary}, Canada}
\IEEEauthorblockA{\IEEEauthorrefmark{1}{Qoherent Inc.}, 
{Toronto, Ontario, Canada}} 
}

\maketitle

\begin{abstract}
Foundation deep learning (DL) models are general models, designed to learn general, robust and adaptable representations of their target modality, enabling finetuning across a range of downstream tasks. These models are pretrained on large, unlabeled datasets using self-supervised learning (SSL). Foundation models have demonstrated better generalization than traditional supervised approaches, a critical requirement for wireless communications where the dynamic environment demands model adaptability. In this work, we propose and demonstrate the effectiveness of a Vision Transformer (ViT) as a \textit{radio foundation model} for spectrogram learning. We introduce a Masked Spectrogram Modeling (MSM) approach to pretrain the ViT in a self-supervised fashion. 
We evaluate the ViT-based foundation model on two downstream tasks: Channel State Information (CSI)-based Human Activity sensing and Spectrogram Segmentation. Experimental results demonstrate competitive performance to supervised training while generalizing across diverse domains. Notably, the pretrained ViT model outperforms a four-times larger model that is trained from scratch on the spectrogram segmentation task, while requiring significantly less training time, and achieves competitive performance on the CSI-based human activity sensing task. This work demonstrates the effectiveness of ViT with MSM for pretraining as a promising technique for scalable foundation model development in future 6G networks.
\end{abstract}


\section{Introduction}
\label{sec:intro}

Foundation models (FMs) are first trained on a large, often unlabeled dataset, allowing them to build broad, adaptable representations that can be finetuned for various downstream tasks. This initial pretraining stage is done using self-supervised learning (SSL), where the model learns underlying patterns and relationships within the data without relying on labeled examples \cite{ssl_survey, ericsson_self-supervised_2022, ssl_wireless}. The model ideally develops a robust understanding of its target modality, which, in our case, is radio spectrograms.

In fields like computer vision and natural language processing, FMs have set new benchmarks \cite{bert, pixelcnn, vis_in_wild, mae}, often surpassing supervised learning models, specifically designed for individual tasks. This is largely due to their ability to generalize: FMs learn flexible and transferable representations that make them better suited to handle variations in data, perform across diverse tasks, and adapt to new contexts. Generalization is especially valuable when labeled data is scarce, as foundation models can perform well with minimal additional labeled samples. 

Deep learning (DL) has demonstrated strong potential when applied to individual wireless tasks, including automatic modulation classification \cite{amc_1}, channel estimation \cite{chan_estim_1}, constellation and waveform design \cite{waveform_1}, among others. However, these models are highly specialized, and there are concerns about their ability to generalize effectively in real-world scenarios. Wireless signals are subject to time-varying impairments, and the communication environment is constantly changing, which can degrade a DL model’s performance if it fails to adapt. Introducing the concept of FMs for wireless can potentially overcome these limitations \cite{llm_phy, llm_telecom}. 

We propose FMs for wireless signals as a solution to address these challenges. By capturing over-the-air radio signals and pretraining FMs through SSL, there is no need for labeled data. Additionally, these pretrained models can then serve as backbones for multiple tasks, reducing computational costs. Most importantly, FMs are expected to achieve better generalization by leveraging their broad, transferable representations, making them well-suited to handle diverse and dynamic wireless environments.
The primary contributions of our paper are:
\begin{itemize} 
\item We propose and demonstrate the effectiveness of a Vision Transformer (ViT) as a radio foundation model for spectrogram learning. Adopting ViT as the FM offers enhanced flexibility, particularly in handling variable input sequences, and increased scalability, as training and evaluation can be parallelized. ViT also captures long-term dependencies through its attention mechanisms.
\item We introduce a Masked Spectrogram Modeling (MSM) approach to pretrain the ViT in a self-supervised fashion, and thoroughly evaluate key design considerations of the masking procedure and transformer size on performance.  
\item By finetuning across two downstream tasks, we demonstrate that the ViT radio FM effectively learns features that generalize across diverse domains, achieving competitive—or even superior—performance with 4x smaller model sizes compared to baselines. 
\item We demonstrate the effectiveness of the proposed foundation model by utilizing a real-world dataset that is captured over-the-air in a software-defined radio testbed.
Upon acceptance, the datasets and code will be publicly available to
encourage further research within the community on FM for wireless.
\end{itemize}

The remainder of the paper is structured as follows: 
Section \ref{sec:datasets} presents the datasets utilized for pretraining the foundation model, and for the CSI-based human activity sensing and spectrogram segmentation tasks.
Section \ref{sec:methods} outlines the ViT architecture and algorithm of the self-supervised foundation model.
Section \ref{sec:results} presents numerical experiments conducted to evaluate the proposed methodology. Finally, section \ref{sec:conclusion} concludes the paper.

\section{Testbed and Datasets}
\label{sec:datasets}

We use three datasets in this paper. The first, the Real-time Radio Dataset (RRD), consists of over-the-air radio recordings captured in real-time with a software-defined radio (SDR) test bed built using PlutoSDRs. The second, the Human Sensing Dataset (HSD), utilizes Wi-Fi channel state information (CSI) to detect human activity in an indoor environment. The third dataset, the Segmentation Dataset (SD), simulates 5G New Radio (NR) and LTE transmissions in neighboring frequency bands.
\subsection{Real-time Radio Dataset (RRD)}

The RRD dataset consists of recordings of IQ samples, representing both in-phase (I) and quadrature (Q) components of the RF signal. Each recording is captured with a center frequency (ranging from $2.4$ to $2.65$ GHz), sampling frequency (between $10$ MHz and $60$ MHz), and duration, typically averaging around $100$ ms. Data collection took place in downtown Toronto, Canada, resulting in $240$ recordings, which cover approximately $24$ seconds of RF activity. This dataset is used for initial model pretraining.\\
\textbf{Spectrogram Computation.} We create spectrograms from IQ recordings through the following steps: 1) Divide each recording into non-overlapping $16$ ms segments; 2) Compute the spectrogram for each segment using the short-time Fourier transform (STFT); 3) Resize each spectrogram to a $224 \times 224$ shape; 4) Convert the spectrogram to log scale; 5) Normalize and standardize using dataset-wide statistics.

The dataset parameters are summarized in Table \ref{tab:rrd_params}. Learning performance is generally robust to these specific parameter choices, which are selected to balance computational efficiency and preserve information-rich content.

\subsection{Human Activity CSI-Based Sensing Dataset (HSD)}

\begin{figure}[h!]
    \centering
    \includegraphics[width=0.9\linewidth]{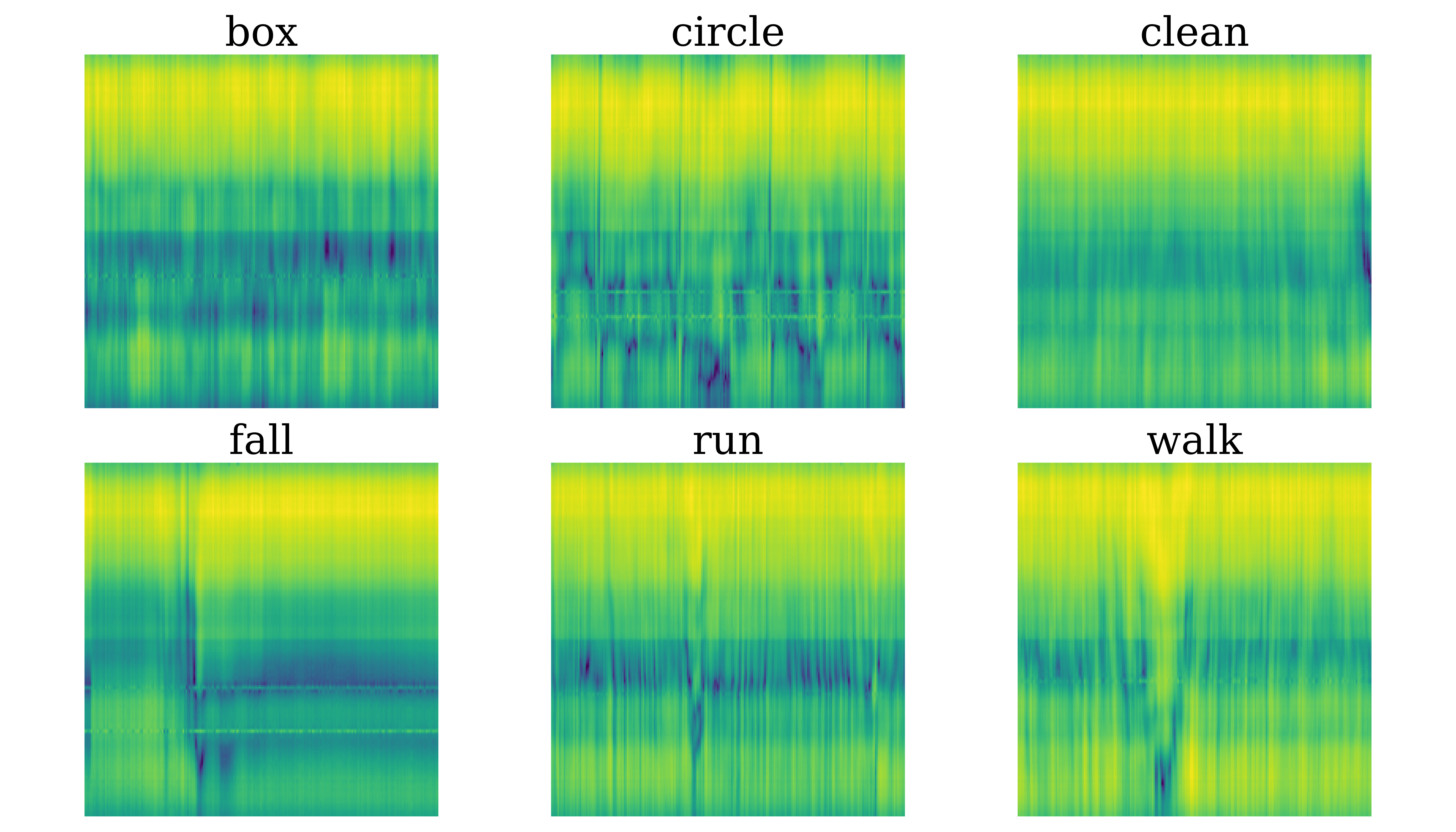}
    \caption{A sample from each class of the HSD dataset. Only the CSI of the first antenna is plotted.}
    \label{fig:csi_sensing_sample}
\end{figure}

The HSD dataset contains CSI measurements for six human activities: running, walking, falling, boxing, arm circling, and floor cleaning \cite{csi_sensing}. Each subject performs these activities between a pair of Wi-Fi access points, each equipped with three antennas. CSI is measured for each activity, across $114$ subcarriers and $3$ channels (one per antenna) over $2000$ samples at a $500$ Hz rate. Each recording is thus a 3D tensor of shape $3 \times 114 \times 2000$, paired with its activity label. The CSI is processed by resizing each recording to a shape of $3 \times 224 \times 224$. Then, each channel is normalized and standardized using dataset-wide statistics. A sample from each class of the dataset is illustrated in Figure \ref{fig:csi_sensing_sample} where the horizontal axis is time and the vertical axis is frequency.
\begin{table}[h!]
    \caption{RRD Dataset Generation Parameters}
    \renewcommand{\arraystretch}{1.2}
    \setlength{\arrayrulewidth}{0.3mm} 
    \setlength{\tabcolsep}{12pt}
    \centering
    \begin{tabular}{ccc}
    \multicolumn{2}{c}{\textbf{Parameters}}  & \textbf{Value} \\ \hline
    \multicolumn{1}{c}{\multirow{4}{*}{\begin{tabular}[c]{@{}c@{}}\textbf{STFT} \\ \textbf{Parameters}\end{tabular}}} & FFT Size  &  $1024$ \\ 
    \multicolumn{1}{c}{} & Window Function &  Hanning \\
    \multicolumn{1}{c}{} & Window Size & $512$ \\
    \multicolumn{1}{c}{} & Hop Size & $512$ \\ 
    \noalign{\hrule height 0.25mm}
    \multicolumn{1}{c}{\multirow{2}{*}{\begin{tabular}[c]{@{}c@{}}\textbf{Slicing} \\ \textbf{Parameters}\end{tabular}}} & Duration  & $16$ ms \\
    \multicolumn{1}{c}{} & Resizing Shape  & $(224,\ 224)$ \\ 
    \end{tabular}
    \vspace{5pt}
    \label{tab:rrd_params}
\end{table}
\begin{figure}[h!]
    \centering
    \includegraphics[width=0.9\linewidth]{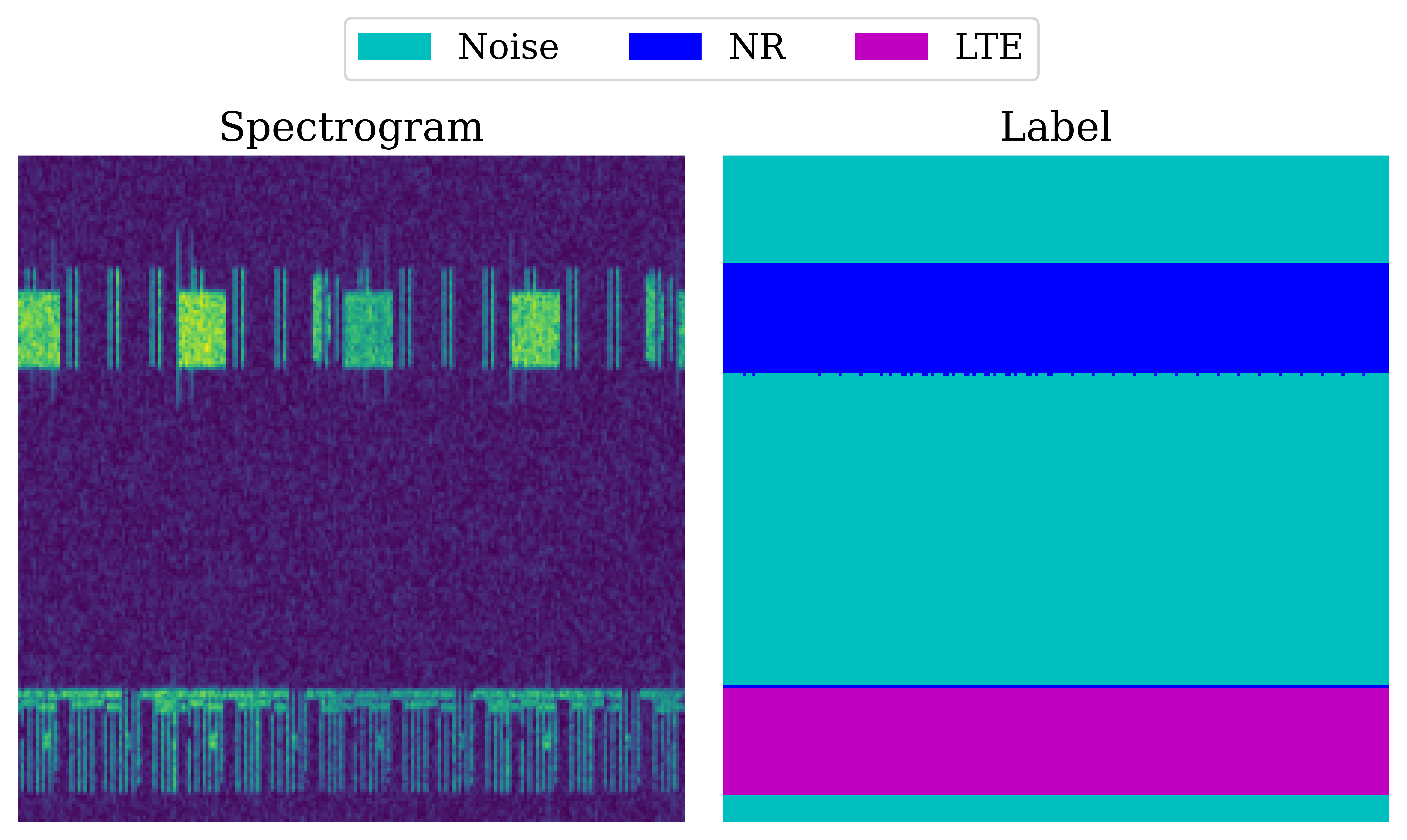}
    \caption{A spectrogram and its segmentation from the SD dataset.}
    \label{fig:spect_label_segm}
\end{figure}

\subsection{NR-LTE Segmentation Dataset (SD)}

The SD dataset is created by generating NR and LTE signals, each transmitted through its respective wireless channel in adjacent, non-overlapping bands. We use the Matlab Communication Toolbox for signal generation, following the guidelines in \cite{5g_segmentation_matlab}.

A spectrogram of the NR-LTE signal mixture is computed and resized to $224 \times 224$. A corresponding label image is also created, marking NR signals as $1$, LTE signals as $2$, and noise as $0$. For more details about data generation, refer to \cite{gc_msm}. A sample is illustrated in Figure \ref{fig:spect_label_segm} where the horizontal axis represents time and the vertical axis represents frequency.


\section{Vision Transformer Foundation Model for Spectrogram Learning}
\label{sec:methods}

\subsection{Masked Spectrogram Modeling (MSM)}

We introduce the Masked Spectrogram Modeling (MSM) approach using Vision Transformers (ViT). In this method, we divide each spectrogram image into $p \times p$ patches and randomly sample a subset of these patches using a uniform distribution. The goal is to reconstruct the missing patches from only the sampled subset, while the remaining patches—effectively the masked patches—are excluded. This approach is illustrated in Figure \ref{fig:mae_setup}. 
While this approach resembles a traditional auto-encoder, a key difference is that the model is trained to reconstruct the masked patches only, rather than the full set.

We employ high masking ratios (e.g., $80\%$) as in \cite{mae} to reduce redundancy and make reconstruction more challenging. This forces the model to rely less on extrapolation from visible patches, 
effectively avoiding learning features that are more local, and instead emphasizing general characteristics that contribute to the overall representation of the spectrogram, its underlying structure and statistical patterns. 

This approach offers several advantages: masking a large portion of the spectrogram and only processing the visible patches makes pretraining more efficient. This method requires no labeled data, recordings can be captured over-the-air using software-defined radios (as we have done with the RRD dataset) and fed into the model directly, making large-scale pretraining more feasible.

\subsection{Spectrogram Masked ViT Autoencoder}

\begin{figure}[h!]
    \centering
    \includegraphics[trim=25 25 25 25, width=0.9\linewidth, keepaspectratio]{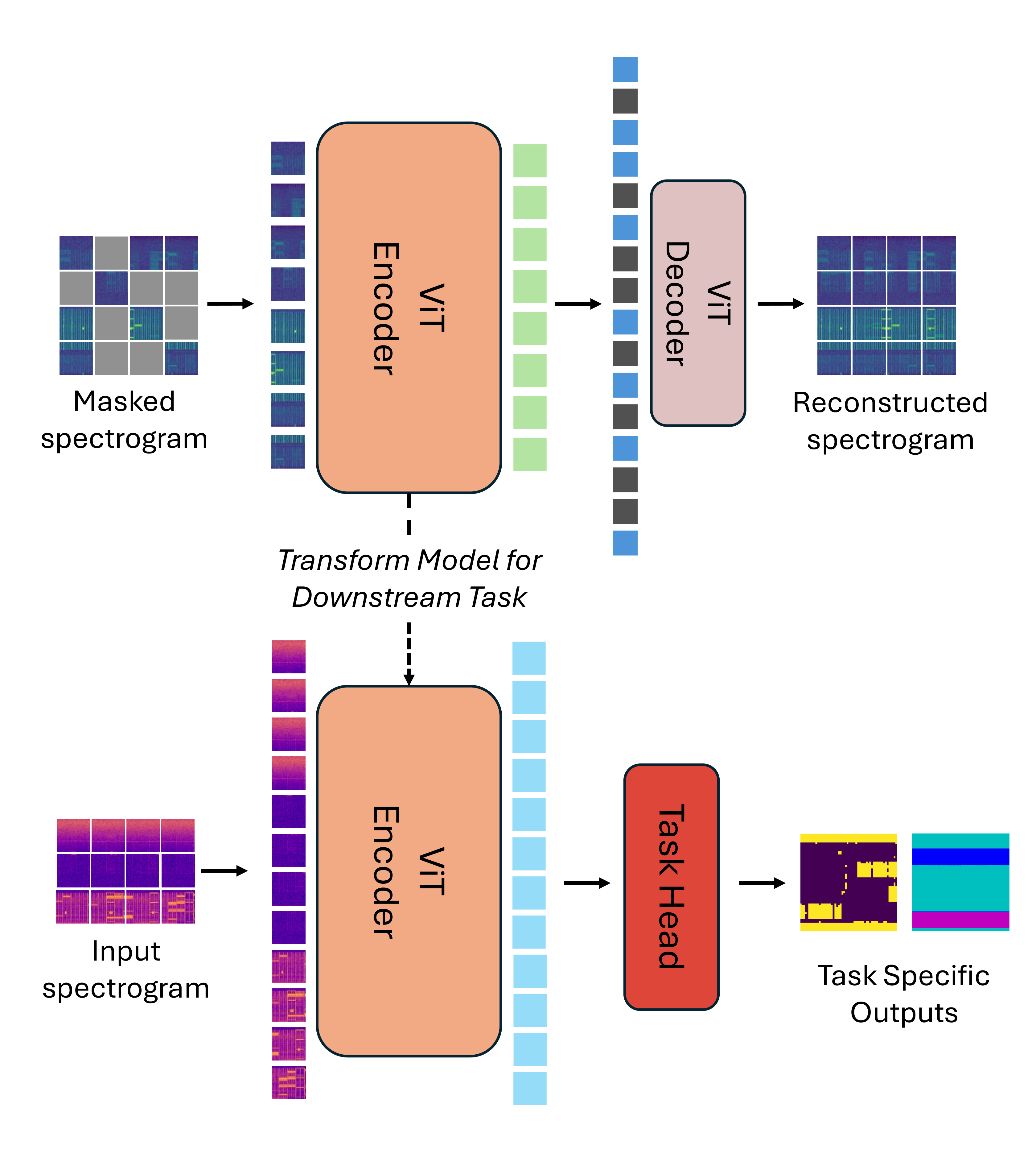}
    \caption{Proposed ViT Foundation Model for Radio Spectrograms.}
    \label{fig:mae_setup}
\end{figure}

As shown in Figure \ref{fig:mae_setup}, we use an encoder-decoder architecture based on a ViT masked autoencoder \cite{vit, mae}. This design is asymmetric in several respects. The encoder processes the visible patches outputting feature tokens, and the decoder handles the feature and mask tokens.
The decoder reconstructs the original spectrogram by attending to the feature tokens provided by the encoder. 
\begin{algorithm}[h!]
    \SetKwInOut{Input}{Input}
    \SetKwInOut{Output}{Output}
    \Input{spectrogram dataset $\mathcal{D}$, initial model $\mathcal{M}$,  patch size $p$, mask ratio $\gamma$}
    \Output{foundation model $\mathcal{B}$}
    $\mathcal{B} \gets $ encoder of the ViT model $\mathcal{M}$\\
    \Repeat{\normalfont convergence is reached or another stopping condition is met}{
        \ForEach{\textit{spect\_sample} in $\mathcal{D}$}{ 
            \textit{patches} $\gets$ {\resizebox{\width}{1.2\height}P}ATCHIFY\big(\textit{spect\_sample}, $p$\big) \\
            \textit{visible\_patches} $\gets$ {\resizebox{\width}{1.2\height}S}AMPLE\big(\textit{patches}, $\gamma$\big) \\ 
            \textit{encoder\_in} $\gets \mathcal{M}\cdot \text{{\resizebox{\width}{1.2\height}E}MBED}\text{\big(}\textit{visible\_patches}\text{\big)}$\\ 
            \textit{encoder\_in} $\gets$ {\resizebox{\width}{1.2\height}P}OS\_{\resizebox{\width}{1.2\height}E}MBED\big(\textit{encoder\_in}\big)\\
            \textit{encoder\_out} $\gets \mathcal{M}\cdot \text{{\resizebox{\width}{1.2\height}E}NCODE}\text{\big(}\textit{encoder\_in}\text{\big)}$ \\
            \textit{decoder\_in} $\gets \mathcal{M}\cdot \text{{\resizebox{\width}{1.2\height}D}ECODER\_{\resizebox{\width}{1.2\height}E}MBED}\text{\big(}\textit{encoder\_out}\text{\big)}$ \\
            \textit{decoder\_in} $\gets$ {\resizebox{\width}{1.2\height}A}PPEND\_{\resizebox{\width}{1.2\height}R}EORDER\big(\textit{decoder\_in}\big) \\
            \textit{decoder\_in} $\gets$ {\resizebox{\width}{1.2\height}P}OS\_{\resizebox{\width}{1.2\height}E}MBED\big(\textit{decoder\_in}\big) \\
            \textit{decoder\_out} $\gets \mathcal{M}\cdot \text{{\resizebox{\width}{1.2\height}D}ECODE}\text{\big(}\textit{decoder\_in}\text{\big)}$ \\
            \textit{recon\_patches} $\gets$ {\resizebox{\width}{1.2\height}U}NPATCHIFY\big(\textit{decoder\_out}\big)\\
            \textit{loss} $\gets \mathcal{L}_{\text{MSM}}\text{\big(}\textit{recon\_patches}, \textit{visible\_patches}\text{\big)}$ as per equation \eqref{eq:loss_masked}\\
            {\resizebox{\width}{1.2\height}B}ACKWARD\big($\mathcal{M}$,\ \textit{loss}\big)\\
        }
    }
    \caption{Masked Spectrogram Modeling with ViT}
    \label{alg:msm} 
\end{algorithm}
    
\begin{table}[h!]
    \centering
    \setlength{\arrayrulewidth}{0.4mm} 
    \renewcommand{\arraystretch}{1.15}
    \begin{tabular}{lp{0.55\linewidth}}
        \textbf{Function} & \textbf{Description} \\
        \hline
        \\[-1em]
        {\resizebox{\width}{1.2\height}P}ATCHIFY & Splits the input spectrogram into smaller patches of a specified size $p \times p$. \\
        {\resizebox{\width}{1.2\height}S}AMPLE & Selects a subset of patches based on the mask ratio $\gamma$. \\
        {\resizebox{\width}{1.2\height}E}MBED* & Prepare the visible patches for the encoder by mapping them to its embedding space. \\
        {\resizebox{\width}{1.2\height}P}OS\_{\resizebox{\width}{1.2\height}E}MBED & Adds sinusoidal positional embeddings to its input.\\ 
        {\resizebox{\width}{1.2\height}E}NCODE* & Processes the embedded patches through the encoder transformer blocks. \\
        {\resizebox{\width}{1.2\height}D}ECODER\_{\resizebox{\width}{1.2\height}E}MBED* & Prepares the encoder output for the decoder by mapping it to the decoder embedding space. \\
        {\resizebox{\width}{1.2\height}A}PPEND\_{\resizebox{\width}{1.2\height}R}EORDER & Reorders tokens and inserts mask tokens to restore the original time-frequency order. \\
        {\resizebox{\width}{1.2\height}D}ECODE* & Processes the reordered tokens through the decoder transformer blocks, producing the reconstructed patches. \\
        {\resizebox{\width}{1.2\height}U}NPATCHIFY & Combines reconstructed patches back into a complete spectrogram. \\
        {\resizebox{\width}{1.2\height}B}ACKWARD & Computes gradients and updates the model using Backpropagation \\
    \end{tabular}
    \caption{Explanation of Functions in Algorithm \ref{alg:msm} (* denotes functions called through the model)}
    \label{tab:function_explanations}
    \vspace{-0.5cm}
\end{table}
Masked tokens are learnable embeddings which are positioned in the original locations of the masked patches (i.e., not inputted to the encoder). 

The encoder is larger than the decoder in terms of capacity, it performs the majority of the computation. As a result, the encoder can function independently as a feature extractor, while the decoder can be discarded. The approach is detailed in Algorithm \ref{alg:msm}. In the following, we provide the high-level details of the ViT architecture.\\
\textbf{Encoder.} Each input patch is embedded using a linear projection, and sinusoidal positional embeddings are added to create a token. The purpose of the positional embeddings is to indicate the order, as transformers lack a built-in ordering mechanism. The tokens are then processed through a series of transformer blocks, producing the output feature tokens.\\
\textbf{Decoder.} At the decoder, a linear projection is applied to match the feature token dimension to the decoder embedding dimension. The original time-frequency ordering of the resulting tokens is restored with mask tokens inserted in place of the masked patches. Sinusoidal positional embeddings are added as well. The sequence of tokens is then processed by a series of transformer blocks and the output is the reconstructed spectrogram.\\ 
By processing only a subset of the tokens using the larger encoder and handling the full set with the smaller decoder, this design enables the training of much larger models without extensive computational resources.\\
\textbf{Objective.} We train the model in a self-supervised way to reconstruct the masked patches. The loss function $\mathcal{L}_{\text{MSM}}$ of MSM task can be written as:
\begin{equation}
    \label{eq:loss_masked}
    \mathcal{L}_{\text{MSM}} = \dfrac{1}{NM} \sum_{n=1}^{N} \sum_{ij} \left\|\myvec\left(\mathbf{X}_{ij}^{(n)}\right) - \myvec\left(\hat{\mathbf{X}}_{ij}^{(n)}\right)\right\|_2^2 \mathbb{I}_{\text{mask}}(n, i, j) 
\end{equation}
where $N$ is the batch size, $M$ is the total number of patches,  $\mathbf{X}_{ij}^{(n)} \in \mathbb{R}^{p \times p}$ is the input patch at position $(i, j)$ in sample $n$, and $\hat{\mathbf{X}}_{ij}^{(n)} \in \mathbb{R}^{p\times p}$ denotes the reconstructed patch at position $(i, j)$ for sample $n$. The vectorization operation $\myvec$ flattens each patch into a vector, $\|\cdot\|_2$ is the $L_2$ norm and $\mathbb{I}_{\text{masked}}(n, i, j)$ is an indicator function that outputs $1$ if patch $(i, j)$ in sample $n$ was masked and $0$ otherwise. 

The encoder of the self-supervised pretrained ViT masked autoencoder serves as our \textit{radio foundation model} which can be finetuned for downstream tasks. We finetune for two downstream tasks: CSI-based human activity sensing and spectrogram segmentation, introduced next.

\subsection{CSI-based Human Activity Sensing}

The task is to classify CSI measurements into one of six distinct human activity classes. We utilize the ViT encoder from the pretrained model as a feature extractor, adding a linear layer as a classification head on top. The ViT encoder is entirely frozen, only the linear classifier is finetuned on the dataset. The pretrained model was originally trained on single-channel spectrograms, whereas here the input is a three-channel tensor representing the CSI. To accommodate this difference, the positional embeddings are modified to align with the new input, while the remainder of the encoder remains unchanged. The CSI data is divided into patches in the same manner as the spectrograms. The model outputs a softmax probability vector, and the loss function is the label smoothing cross-entropy, defined as:
\begin{equation}
    \label{eq:loss_sensing}
    \mathcal{L}_{\text{HSD}} = -\dfrac{1}{N} \sum_{n=1}^N \sum_{i=1}^{C} \left(y_{i}^{(n)} \cdot (1 - \alpha) + \dfrac{\alpha}{C}\right) \cdot \log\left(\hat{y}_{i}^{(n)}\right)
\end{equation}
where $N$ is the batch size, $C = 6$ is the number of classes, $y_i^{(n)} \in [0, 1]$ is the true label for class $i$ (either $0$ or $1$ for sample $n$), $\hat{y}_i^{(n)} \in [0, 1]$ is the model’s predicted probability for class $i$, and $\alpha \in (0, 1)$ is the smoothing factor. Unlike traditional cross-entropy, label smoothing distributes a small probability to incorrect labels, preventing the model from becoming overly confident which enhances generalization. The degree of smoothing is determined by $\alpha$.

\subsection{Spectrogram Segmentation}

The task is to segment the input spectrogram into three classes: noise, NR signal, and LTE signal. We use the pretrained ViT encoder as a feature extractor, adding two standard transformer decoder blocks on top as a segmentation head. The ViT encoder is kept frozen, and only the decoder is finetuned. Since the input is a spectrogram, no modifications are made to the positional embeddings. The model’s output is a 3D tensor providing a probability distribution for each pixel in the segmented spectrogram. We use label smoothing cross-entropy as the loss function, which is defined as follows:
\begin{equation}
    \label{eq:loss_segmentation}
    \mathcal{L}_{\text{SG}} = - \dfrac{1}{NM} \sum_{n=1}^{N} \sum_{k=1}^{C} \sum_{ij}  \left(y_{ijk}^{(n)} \cdot (1 - \alpha) + \dfrac{\alpha}{C}\right) \cdot \log\left(\hat{y}_{ijk}^{(n)}\right)
\end{equation}
Here, $M$ is the total number of pixels in the segmented image, $C=3$ is the number of classes, $y_{ijk}^{(n)} \in [0, 1]$ is the correct label at pixel $(i, j)$ for class $k$ in sample $n$, $\hat{y}_{ijk}^{(n)} \in [0, 1]$ is the predicted probability at pixel $(i, j)$ for class $k$, and $\alpha$ is the smoothing factor.

\section{Results and Discussion}
\label{sec:results}

We perform self-supervised pretraining with masking on the RRD dataset, then evaluate the learned representations by finetuning. For finetuning, the decoder is discarded, and the frozen ViT encoder serves as a feature extractor, with only the task-specific head updated. No masking is done during finetuning. Three models are pretrained: ViT-S (small), ViT-M (medium), and ViT-L (large), with details provided in Table \ref{tab:model_params}. Here, different sizes refer to the encoder, while the decoder remains largely unchanged. First, we evaluate the models' reconstruction performance across various masking ratios, followed by assessing generalization capabilities on the CSI Sensing and Segmentation datasets.

\begin{table*}[h!]
    \caption{Pretrained ViT Models}
    \renewcommand{\arraystretch}{1.15}
    \setlength{\arrayrulewidth}{0.4mm} 
    \centering
    \begin{tabular}{cccccccccc}
    \multicolumn{1}{c|}{\multirow{2}{*}{Model}} & \multicolumn{5}{c|}{Encoder}                                                        & \multicolumn{4}{c}{Decoder}                       \\
    \multicolumn{1}{c|}{}  & patch size & embed dim\textsuperscript{*} & depth\textsuperscript{†} & attn. heads & \multicolumn{1}{c|}{\# params (M)} & embed dim\textsuperscript{*} & depth\textsuperscript{†} & attn. heads & \# params (M) \\ \hline
    ViT-S & $16$ & $512$ & $12$ & $8$ & $38$ & $256$ & $8$ & $16$ & $7$  \\
    ViT-M & $16$ & $768$ & $12$ & $12$ & $85$ & $512$ & $8$ & $16$ & $26$ \\
    ViT-L & $16$ & $1024$ & $24$ & $16$ & $302$ & $512$ & $8$ & $16$ & $26$ \\          
    \end{tabular}
    \vspace{5pt}
    \caption*{\scriptsize \centering * \textit{embed dim} is the embedding dimension which is also known as the transformer width.\protect\linebreak† \textit{depth} is the number of transformer blocks.}
    \label{tab:model_params}
    \vspace{-0.25cm}
\end{table*}

\subsection{Reconstruction Performance}

\begin{figure}[h!]
    \centering
    \includegraphics[trim=0 0 0 25, width=0.9\linewidth, keepaspectratio]{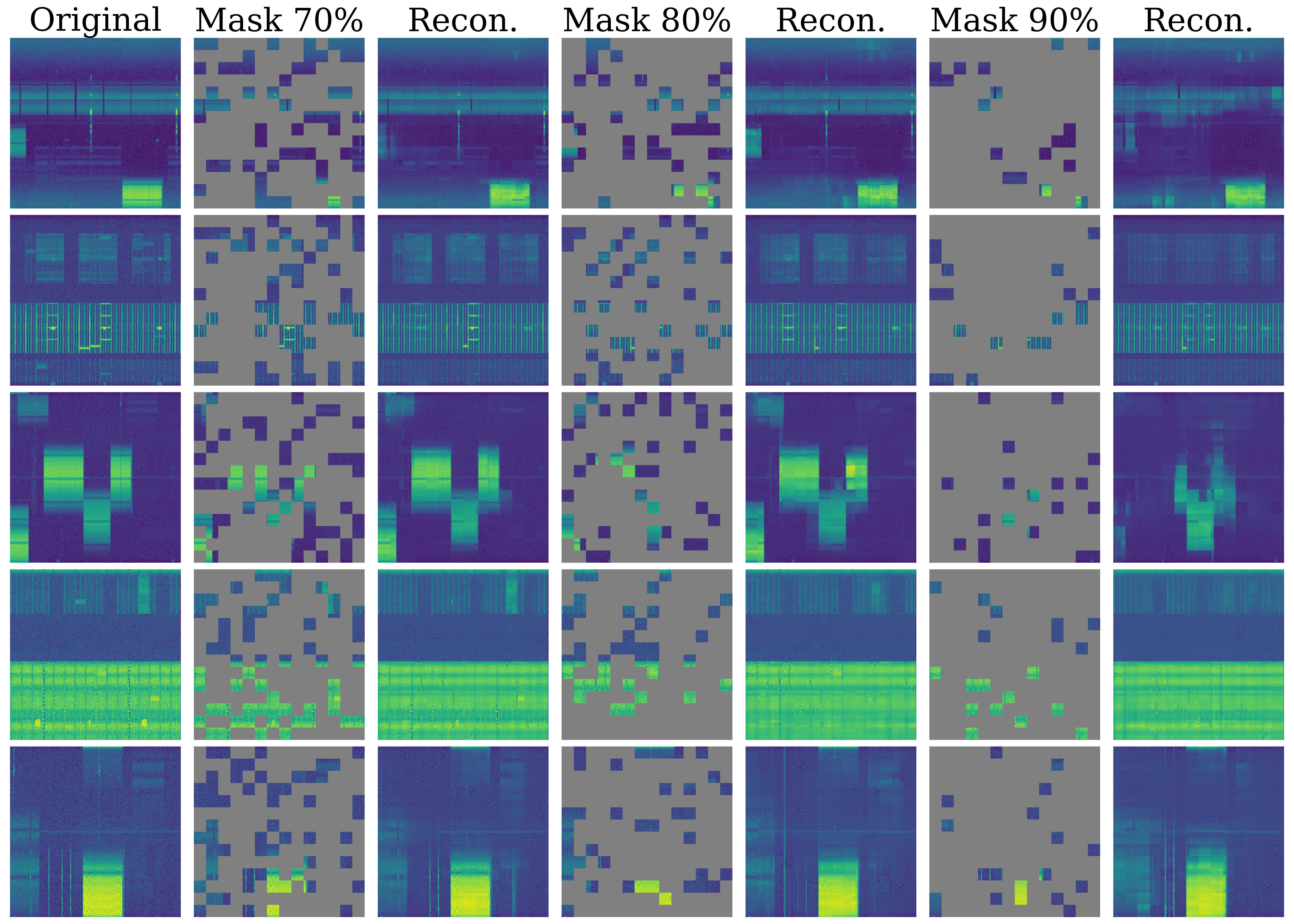}
    \caption{Reconstruction results of ViT-M at various masking ratios pretrained with a $75\%$ masking ratio.} 
    \label{fig:reconstructed_images}
\end{figure}

First, we showcase reconstruction examples for ViT-M from which it is clear that the model exhibits strong performance. This is illustrated in Figure \ref{fig:reconstructed_images}. Each row shows the original spectrogram on the left, followed by the masked spectrogram and the corresponding model's reconstruction for different masking ratios. The reconstructed spectrograms closely match the originals, with reasonable differences.
\begin{figure}[h!]
    \centering
    \includegraphics[width=0.9\linewidth, keepaspectratio]{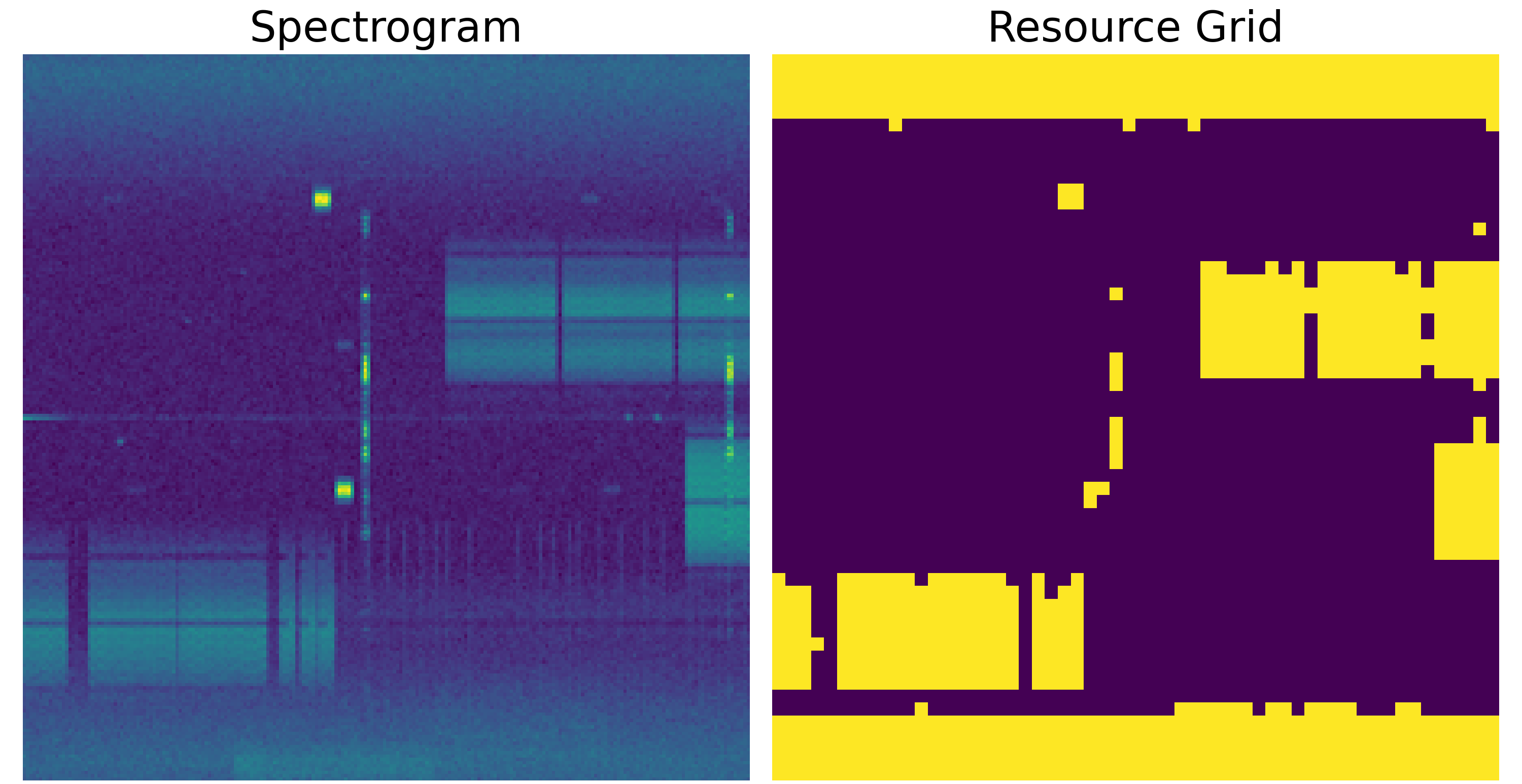}
    \caption{A spectrogram and its corresponding resource grid using a pooling filter of size 4.}
    \label{fig:spect_grid}
\end{figure}
\begin{figure}[h!]
    \centering
    \includegraphics[width=0.9\linewidth, keepaspectratio]{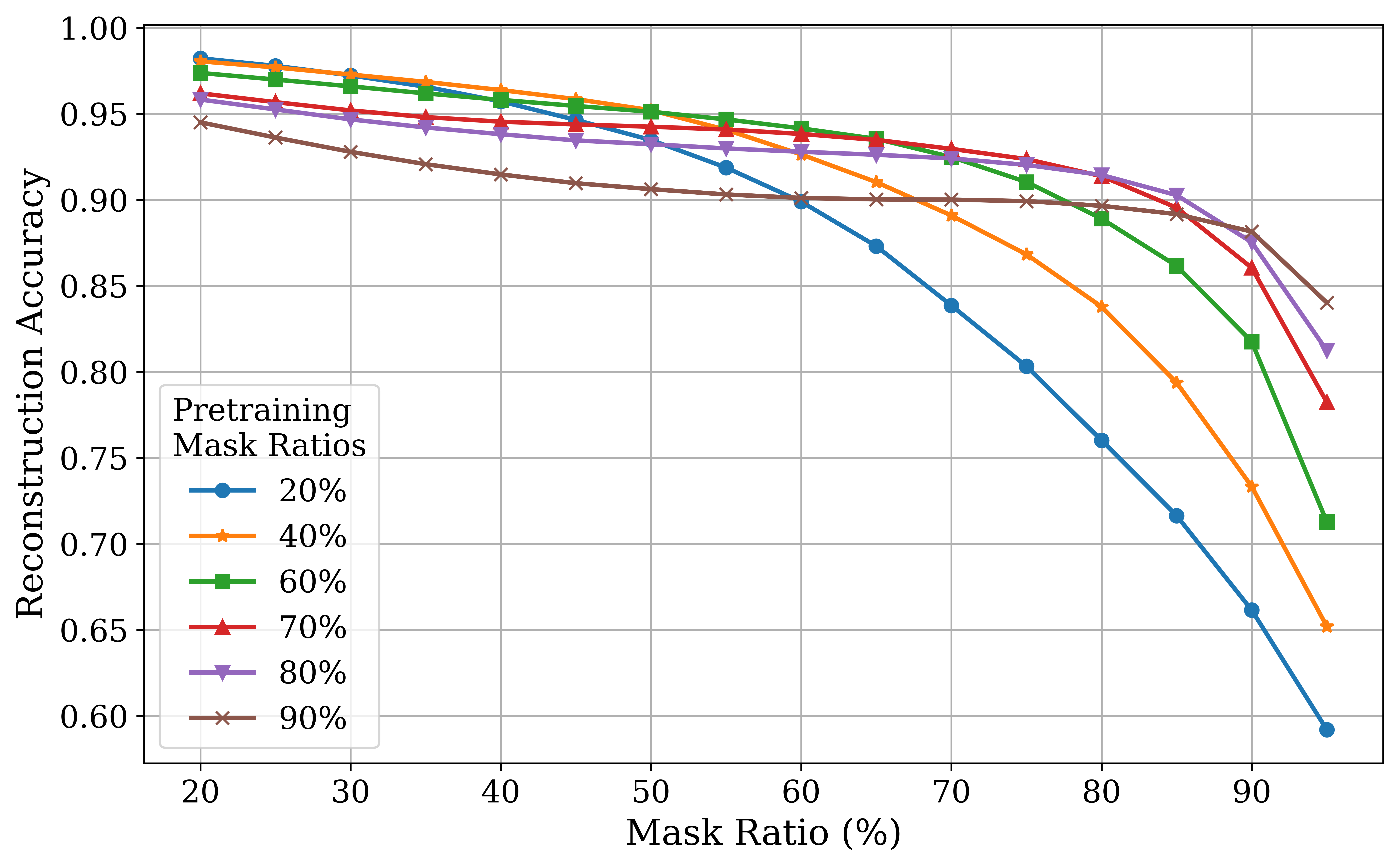}
    \caption{Reconstruction Accuracy vs Mask ratio of ViT-S pretrained at various masking ratios.}
    \label{fig:recon_gen}
    \vspace{-0.25cm}
\end{figure}
To evaluate the reconstruction capability of the models, we need a robust reconstruction accuracy metric. Relying solely on visual comparison is not enough. Hence, we transform each spectrogram into a resource grid composed of resource blocks. To transform the spectrogram. average pooling is first applied without overlap between pooled patches (i.e., stride equal to the kernel size).  
A threshold, $\delta$, is then applied to the pooled grid to binarize it, designating vacant resource blocks as $0$ and occupied ones as $1$. The threshold $\delta$ is determined empirically by the formula:
\begin{equation}
\label{eq:threshold}
\delta = \mu + 0.5 \times \sigma
\end{equation}
where $\mu$ and $\sigma$ represent the mean and standard deviation of the spectrogram, respectively. A sample for the transformation is illustrated in Figure \ref{fig:spect_grid}.
We then evaluate the models' reconstruction performance across various masking ratios, including but not limited to those used during pretraining. A model’s robustness is measured by its ability to maintain strong reconstruction performance even when applied to masking ratios it was not trained on. Although higher masking ratios increase reconstruction difficulty, the model is expected avoid collapse. As illustrated in Figure \ref{fig:recon_gen}, pretrained models with higher masking ratios maintain their strong performance when dealing with different masking ratios. Similar to results for vision and audio, the ideal masking ratio for pretraining is around $70\%$ to $80\%$. Hence, we only finetune the models pretrained with these masking ratios.

\subsection{Finetuning Performance}

\begin{figure}[h!]
    \centering
    \includegraphics[width=\linewidth, keepaspectratio]{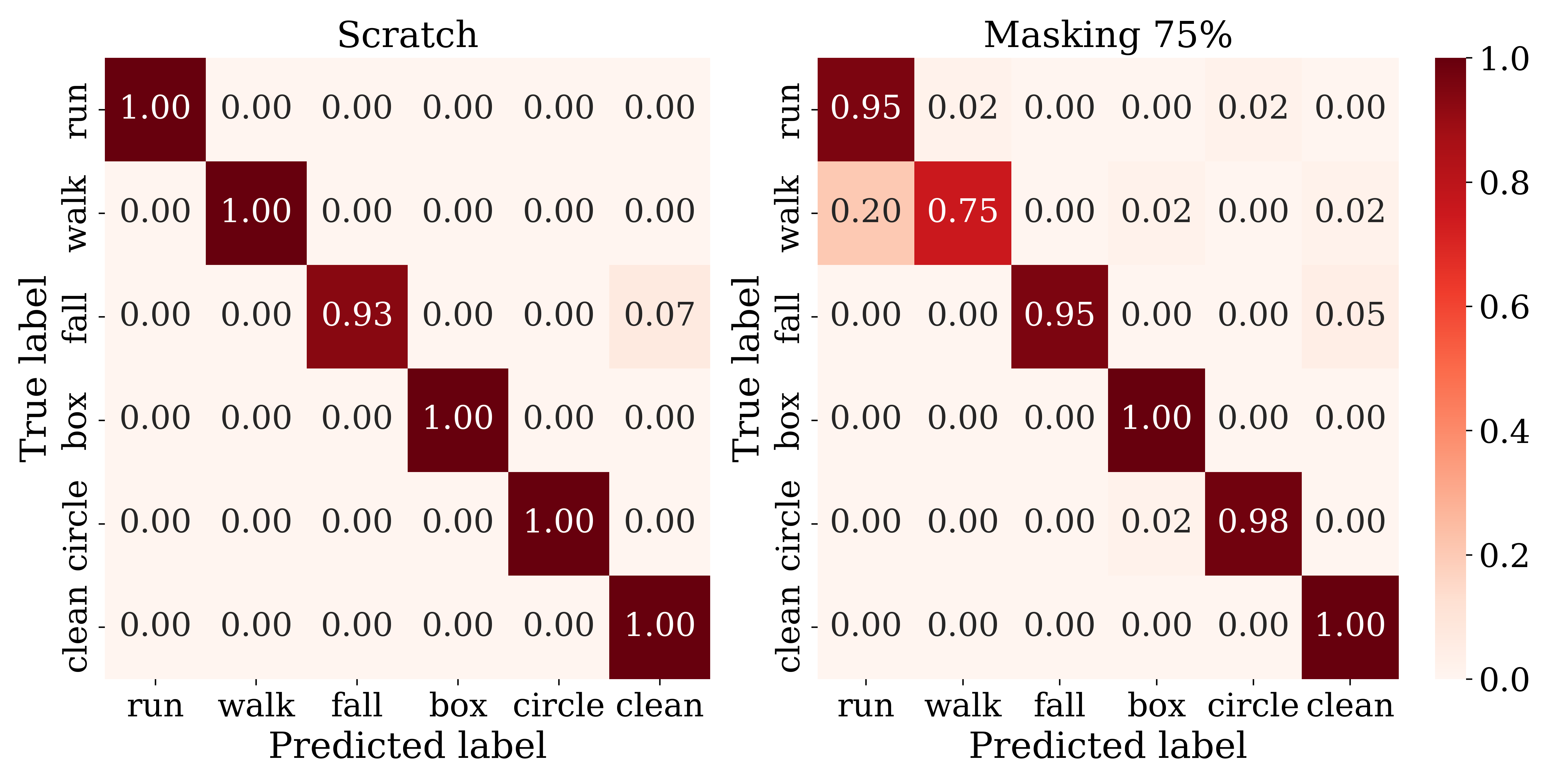}
    \caption{Confusion matrices of ViT-M trained from scratch and pretrained with a $75\%$ masking ratio.}
    \label{fig:conf_mats_medium_sensing}
\end{figure}

To evaluate finetuning performance on the HSD and SD classification and segmentation datasets, we use confusion matrices (per-class accuracy) and overall accuracy. We present finetuning results for the HSD dataset first, followed by the SD dataset. For both datasets, we use the pretrained ViT encoder as a feature extractor which is kept frozen, and finetune the task-specific head. Table \ref{tab:sensing_accuracies} summarizes the accuracy results, including models pretrained at masking ratios of $70\%$, $75\%$, and $80\%$, as well as a baseline model trained from scratch directly on the HSD dataset.
The highest accuracy is achieved by ViT-M trained from scratch, with a $5\%$ accuracy margin compared to the pretrained ViT-M model. We attribute this difference to the inherent distinctions between CSI data and spectrograms, suggesting that more extensive pretraining could reduce this gap. Figure \ref{fig:conf_mats_medium_sensing} displays the confusion matrices for ViT-M pretraining at a $75\%$ masking ratio versus training from scratch. The primary source of accuracy differences lies in the pretrained model's tendency to confuse run and walk, due to their close distribution.

For the SD dataset, Table \ref{tab:segmentation_accuracies} presents the results, including models pretrained with masking ratios of $70\%$, $75\%$, and $80\%$, alongside a baseline model trained from scratch.
\begin{table}[h!]
    \caption{Mean accuracy of ViT finetuned on the HSD dataset, pretrained at masking ratios of $70\%$, $75\%$, and $80\%$. The table also includes results for a model trained from scratch.}
    \renewcommand{\arraystretch}{1.5}
    \setlength{\arrayrulewidth}{0.3mm} 
    \setlength{\tabcolsep}{12pt}
    \centering
    \begin{tabular}{ccccc}
          & \multicolumn{3}{c}{\textbf{Masking ratio} (\%)} & \multirow{2}{*}{\textbf{Scratch}} \\
    \textbf{Model} & $70$ & $75$ & $80$ & \\ \hline
    ViT-S & $90.2$ & $90.9$ & $89.0$ &  $98.1$  \\
    ViT-M & $92.0$ & \cellcolor{verylightgray}$93.9$ & $85.9$ & \cellcolor{verylightgray}$98.9$ \\
    ViT-L & $89.3$ & $88.6$ & $85.6$ & $98.1$                   
    \end{tabular}
    \label{tab:sensing_accuracies}
\end{table}

\begin{table}[h!]
    \caption{Mean accuracy of ViT finetuned on the SD dataset, pretrained at masking ratios of $70\%$, $75\%$, and $80\%$. }
    \renewcommand{\arraystretch}{1.5}
    \setlength{\arrayrulewidth}{0.3mm} 
    \setlength{\tabcolsep}{12pt}
    \centering
    \begin{tabular}{ccccc}
          & \multicolumn{3}{c}{\textbf{Masking ratio} (\%)} & \multirow{2}{*}{\textbf{Scratch}} \\
    \textbf{Model} & $70$ & $75$ & $80$ & \\ \hline
    ViT-S & $97.0$ & $96.8$ & $96.4$ & $97.2$ \\
    ViT-M & \cellcolor{verylightgray}$97.9$ & $97.6$ & $97.5$ & $97.1$ \\
    ViT-L & $97.5$ & $97.3$ & $97.5$ & \cellcolor{verylightgray}$97.7$                   
    \end{tabular}
    \label{tab:segmentation_accuracies}
\end{table}
\begin{figure}[h!]
    \centering
    \includegraphics[width=\linewidth, keepaspectratio]{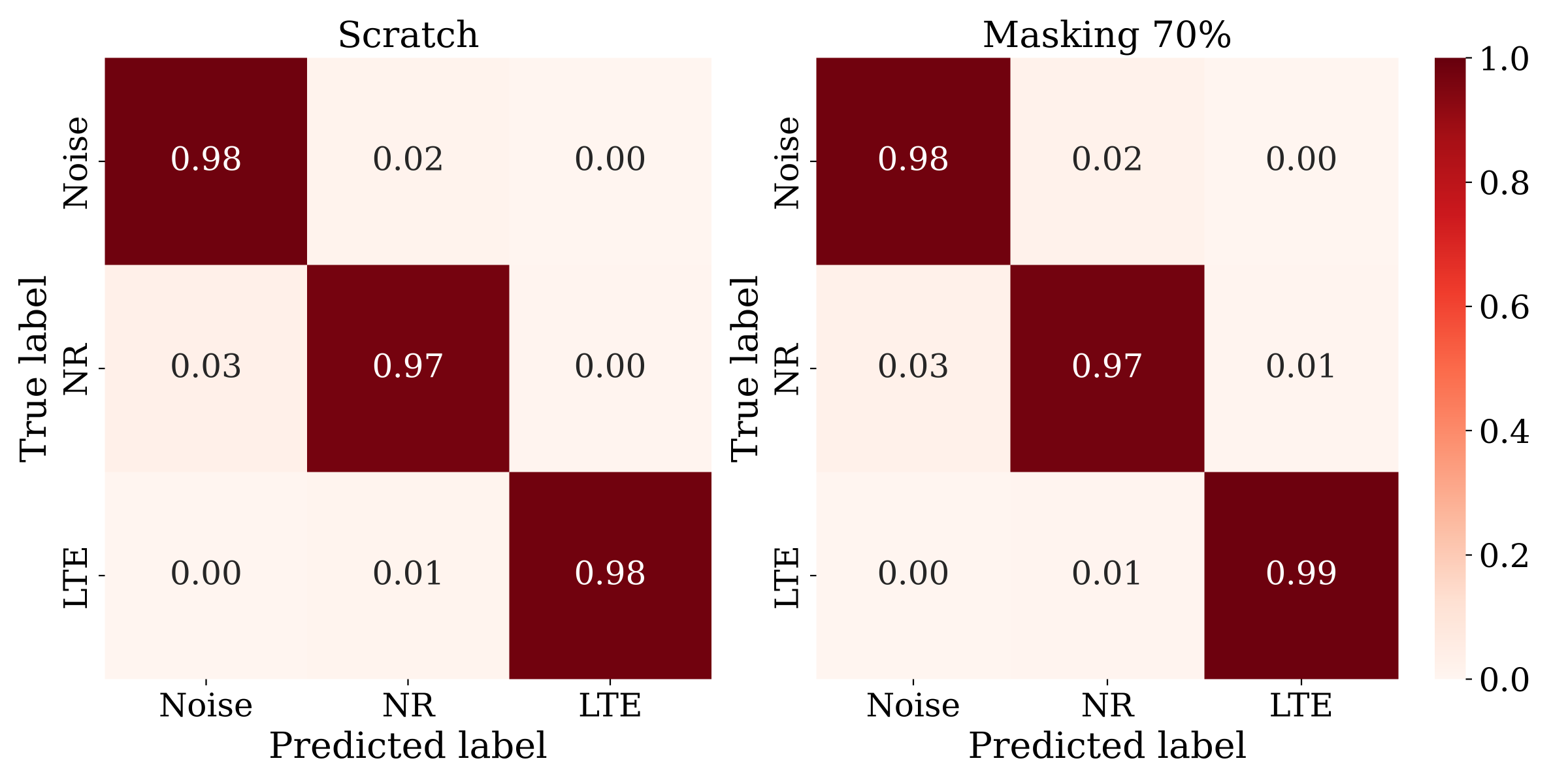}
    \caption{Confusion matrices of ViT-L trained from scratch and ViT-M pretrained with a $70\%$ masking ratio.}
    \label{fig:conf_mats_segm}
\end{figure}
The best model is the pretrained ViT-M with a $70\%$ masking ratio, which slightly outperforms the best scratch-trained model, ViT-L, while being four times smaller. Figure \ref{fig:conf_mats_segm} provides confusion matrices for these models.

\section{Conclusion}
\label{sec:conclusion}

In this paper, we proposed ViT as a \textit{radio foundation model} for spectrogram learning which offers superior modelling capabilities, support for variable-length input sequences and computational efficiency. We also introduce a Masked Spectrogram Modeling (MSM) approach to pretrain the ViT in a self-supervised fashion, and thoroughly evaluate the effects of masking ratios and transformer size on performance.  
Experimental results indicate that the ViT-based model generalizes well to unseen datasets, achieving comparable or superior performance to larger models trained from scratch, while utilizing fewer resources. Notably, the pretrained ViT model surpasses a four-times larger scratch-trained model on the spectrogram segmentation task and achieves competitive performance on the CSI-based human activity sensing task. We believe that this ViT-enabled MSM will enable scalable, large-scale pretraining, fostering the development of robust radio foundation models capable of generalizing across a wide range of tasks.

\bibliography{bibliography.bib}
\bibliographystyle{ieeetr}

\end{document}